\documentclass{osa-article}
\usepackage{multirow}
\journal{boe}
\articletype{Research Article}
\begin{document}

\title{Resolution enhancement and realistic speckle recovery with generative adversarial modeling of micro-optical coherence tomography}

\author{Kaicheng Liang,\authormark{1,*} Xinyu Liu,\authormark{2,3,*} Si Chen,\authormark{2} Jun Xie,\authormark{2} Wei Qing Lee,\authormark{1,4} Linbo Liu,\authormark{2} and Hwee Kuan Lee\authormark{1,3,4,5,6,$\dagger$}}

\address{\authormark{1}Bioinformatics Institute, Agency for Science, Technology and Research (A*STAR), Singapore\\
\authormark{2}School of Electrical and Electronic Engineering, Nanyang Technological University (NTU), Singapore\\
\authormark{3}Singapore Eye Research Institute, Singapore\\
\authormark{4}School of Computing, National University of Singapore (NUS), Singapore\\
\authormark{5}Image and Pervasive Access Lab, CNRS, Singapore\\
\authormark{6}Rehabilitation Research Institute of Singapore, Singapore\\
\authormark{*}Equal contribution\\
\authormark{$\dagger$}leehk@bii.a-star.edu.sg} %% email address is required



\begin{abstract}
A resolution enhancement technique for optical coherence tomography (OCT), based on Generative Adversarial Networks (GANs), was developed and investigated. GANs have been previously used for resolution enhancement of photography and optical microscopy images. We have adapted and improved this technique for OCT image generation. Conditional GANs (cGANs) were trained on a novel set of ultrahigh resolution spectral domain OCT volumes, termed micro-OCT, as the high-resolution ground truth ($\sim$1 $\mu$m isotropic resolution). The ground truth was paired with a low-resolution image obtained by synthetically degrading resolution 4x in one of (1-D) or both axial and lateral axes (2-D). Cross-sectional image (B-scan) volumes obtained from \textit{in vivo} imaging of human labial (lip) tissue and mouse skin were used in separate feasibility experiments. Accuracy of resolution enhancement compared to ground truth was quantified with human perceptual accuracy tests performed by an OCT expert. The GAN loss in the optimization objective, noise injection in both the generator and discriminator models, and multi-scale discrimination were found to be important for achieving realistic speckle appearance in the generated OCT images. The utility of high resolution speckle recovery was illustrated by an example of micro-OCT imaging of blood vessels in lip tissue. Qualitative examples applying the models to image data from outside of the training data distribution, namely human retina and mouse bladder, were also demonstrated, suggesting potential for cross-domain transferability. This preliminary study suggests that deep learning generative models trained on OCT images from high-performance prototype systems may have potential in enhancing lower resolution data from mainstream/commercial systems, thereby bringing cutting-edge technology to the masses at low cost.
\end{abstract}

%%%%%%%%%%%%%%%%%%%%%%%%%%  body  %%%%%%%%%%%%%%%%%%%%%%%%%%
\section{Introduction}
Optical coherence tomography (OCT) is a 3-dimensional optical imaging technique, and has become part of the standard of care in ophthalmology\cite{fujimoto_development_2016} while growing in importance in other clinical specialties such as gastroenterology\cite{gora_endoscopic_2017}. Axial resolution of OCT is governed by the light source bandwidth, while lateral (transverse) resolution is governed by the numerical aperture (NA) of the illumination beam\cite{liu_computational_2017}. Hardware efforts to improve axial and lateral resolution can be complex, requiring high performance lasers and imaging objectives, and dispersion matching in the reference and sample paths. Computational techniques have been employed to overcome these constraints. For axial resolution, dispersion mismatch can be corrected by compensation algorithms to restore resolution to the ideal limit set by the light source; recent studies have proposed methods to surpass that limit\cite{liu_spectral_2015}. For lateral resolution, traditional deconvolution techniques such as the Richardson-Lucy algorithm have been suggested\cite{hojjatoleslami_image_2013}, while physics-based algorithms such as interferometric synthetic aperture microscopy have also been successful\cite{ralston_interferometric_2007}.

Resolution enhancement of images, known as 'computational super-resolution' in the computer vision literature, is a longstanding and rich area of research, beginning almost four decades ago with the proposed use of multiple low-resolution image frames\cite{tsai_multiframe_1984}, a powerful technique that has continued to improve tremendously in recent years\cite{farsiu_fast_2004,farsiu_multiframe_2006,protter_generalizing_2009}. Super-resolution based on single images has since been proposed\cite{zhang_learning_2015}. Super-resolution algorithms for OCT for enhancing low sampling resolution (subsampled) images while also performing denoising (speckle reduction) have also been previously reported\cite{fang_fast_2013,fang_segmentation_2017}, alongside studies of OCT speckle that have proposed statistical models of OCT signal and noise\cite{dubose_statistical_2018}. Deep learning has found most success in image classification and feature detection tasks\cite{lecun_deep_2015}, and has also had a growing impact in computational imaging and inverse problems\cite{barbastathis_use_2019} such as super-resolution, notably in optical microscopy\cite{belthangady_applications_2019} where deep learning has been used to improve image quality\cite{weigert_content-aware_2018}. Generative adversarial networks (GANs)\cite{goodfellow_generative_2014}, an emerging branch of deep learning, has shown promise in a wide range of imaging applications. GANs use two powerful neural networks competing with each other to greatly enhance the quality and realism of machine-generated images, potentially performing better than a single neural network alone or blind techniques without data priors. Conditional GANs (cGANs) are a flavor of GANs that learn to generate a mapping between two domains by training on image pairs, where a `conditional' image in one domain is co-registered with a ground truth image from another domain\cite{isola_image--image_2016}. These techniques have been investigated in photography, microscopy\cite{wang_deep_2019}, as well as OCT studies\cite{ma_speckle_2018,huang_simultaneous_2019}. The latter studies aimed to remove speckle by training on ground truths that were frame-averaged OCT images with a smoothed appearance. This form of denoising can be preferred in some applications when assessing tissue structures or when images have low signal-to-noise ratio. However, speckle can contain important information about tissue scatterers and blood flow. Also, \cite{huang_simultaneous_2019} studied the enhancement of low sampling resolution by synthesizing low-resolution images with subsampling; while important, this is a different mechanism from low optical resolution, which is related to laser bandwidth and laser spot size. 

In this work we explore the hypothesis that cGANs can be used to enhance the optical axial and lateral resolution of OCT images while preserving and improving the detail of speckle content, trained on an ultrahigh resolution OCT ground truth. Using images obtained by micro-OCT\cite{liu_imaging_2011,cui_flexible_2017} with $\sim$1$\mu$m resolution, axial and lateral resolutions were synthetically degraded by windowing/averaging the interference spectra, producing an intrinsically co-registered set of paired low-high resolution data for training. Injection of noise in the cGAN architecture was found to substantially improve the quality of image generation. Comparisons were made between our approach and several previously reported techniques - classical blind deconvolution without deep learning (Richardson-Lucy deconvolution), a state of the art non-adversarial deep learning approach, and a vanilla cGAN with no noise injection. Models were separately trained on two datasets - mouse skin and human lip. Three use cases were investigated - the conversion of 1-dimensional low resolution (axial or lateral) to high resolution, and the conversion of 2-dimensional (axial+lateral) low resolution to high resolution. The 2-D case was further investigated for the realism of the speckle reconstruction, using a perceptual quality test performed by an OCT expert, where our GAN approach was found to perform better than previous techniques. We also report training details and hyperparameter heuristics that are specific to OCT image generation. To illustrate a potential use case of high-resolution speckle recovery, we demonstrate high-resolution imaging of a blood vessel in labial tissue, where the dynamics of small biological particles may be visualized. Lastly, we show qualitative examples of our models performing enhancement on OCT data from outside of the training data distribution, namely human retina and mouse bladder, suggesting potential for cross-domain transferability.

\begin{figure}[htbp]
\centering\includegraphics[width=10cm]{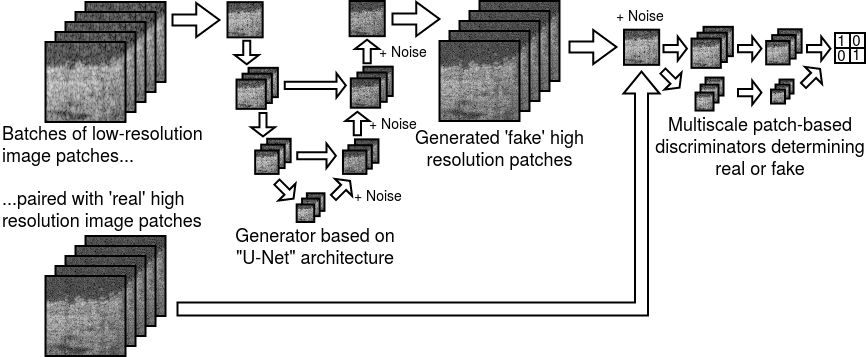}
\caption{Model architecture of conditional GAN (cGAN) for super-resolution. Gaussian noise is injected during upsampling in the generator and at the input to the discriminator, to stabilize training and produce higher quality results.}
\label{fig:archi}
\end{figure}

The paper has the following contributions: 1. we identified critically important modifications to a conventional conditional GAN framework for producing high quality resolution enhancement and realistic speckle generation in OCT images, namely noise injection and multi-scale discrimination, 2. highlighted the utility of extremely high resolution prototype OCT systems such as micro-OCT for the training of resolution-enhancing deep learning tools that can be applied to conventional OCT images, 3. proposed the use of a single human OCT expert to evaluate quality and realism of AI-enhanced images as an alternative to conventional metrics, with the caveat that results are subjective and may not generalize, 4. showed a potential application of high quality speckle recovery towards the study of small biological features/particles using OCT at cellular resolution, 5. showed a potential application of applying AI enhancement tools towards improving resolution of conventional OCT images from commercial systems. Taken together, the paper demonstrates the value of generative and GAN methods, an emerging set of artificial intelligence techniques, when applied to the important field of OCT image analytics and enhancement, potentially delivering broad impact to the large community of OCT users and researchers. 

\section{Materials and methods}

\subsection{micro-OCT image data and pre-processing} 
Images were obtained using a prototype micro-OCT system previously reported\cite{chen_contrast_2019}, with axial scan rate of 60 kHz, axial resolution 1.3 $\mu$m (tissue) and lateral resolution 1.8 $\mu$m. Two datasets were investigated - 10 volumes of mouse skin images from 4 living mice, and 8 volumes of human labial (lip) mucosa images from 2 human subjects, acquired \textit{in vivo} from different regions of tissue by a handheld probe and reported in an earlier publication\cite{chen_contrast_2019}. Models for mouse skin and human lip tissue were trained separately. Images were grouped by volume scans and were allocated to either training or validation data, ensuring that similar B-scans used for training  were not seen during validation. Each volume had dimensions $\sim800\times1000\times500$ ($\sim$500 B-scans per volume). The pixel size was 0.4 $\mu$m (axial) and 0.8 $\mu$m (lateral).  To generate realistic low axial resolution images, a tight Gaussian window with full width at half maximum (FWHM) set to 25\% of the source bandwidth was applied to the raw k-space interference fringe data, degrading the axial resolution to $\sim$5 $\mu$m while preserving the depth dimension. The 4x factor in degradation was selected to produce $\sim$5 $\mu$m, which is in the range of axial resolution for a typical commercial spectral domain OCT system. To generate low lateral resolution images, the fringes were moving-averaged over 6 A-scan lines, corresponding to $\sim$5 $\mu$m in the lateral direction. For 2-D (axial and lateral) low resolution, the fringes were windowed then moving-averaged. Thus the low resolution images were intrinsically co-registered with the high resolution images, with the same pixel dimensions. The images were then cropped to non-overlapping $256\times256$ image patches for model training, with deep low signal regions discarded. The skin volumes were split into 7 volumes for training (28,728 training image patches) and 3 volumes for validation (12,312 validation image patches). The lip mucosa volumes were split into 5 volumes (30,780 training image patches) for training and 3 volumes for validation (18,468 validation image patches). Different models were trained on two versions of data - single frame data, and 3-frame moving averaged data. The 3-frame averaging served as a simple denoising technique that improved the perceptual quality of the images, and is a standard practice in OCT processing when some speckle reduction is preferred.

\subsection{cGAN architecture and training} 

A cGAN architecture was used for the image enhancement deep learning model (Fig. \ref{fig:archi}). In this architecture, two powerful neural networks learn from each other, thereby improving the quality of the outputs. A `generator' neural network learns from paired training data to produce an enhanced image from a `conditional' input image while regularized by a distance metric between the enhanced image and a ground truth image. A generated `fake' or a genuine ground truth image combined with the generator conditional input is fed to a `discriminator' neural network, which learns to discriminate between the genuine and generated images, then returns feedback to the generator. As training of the generator and discriminator models is performed alternately, the two models compete till a theoretical limit where the generated images are indistinguishable from the ground truth, although in practice the generated quality does not necessarily converge to an optimum. This previously reported cGAN design is widely known as `pix2pix'\cite{isola_image--image_2016} and has several open-source skeleton implementations generously made available by the machine learning community\cite{isola_github_2020,linder-noren_github_2020}. We made specific modifications as described below.

\begin{figure}[htbp]
\centering\includegraphics[width=10cm]{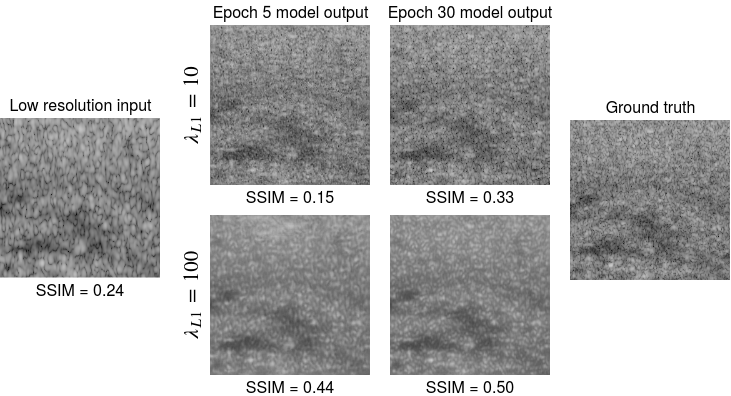}
\caption{Illustrative images showing progress of training and effect of L1 regularization. Image is from validation set. With large regularization, structural similarity (SSIM) values were inflated despite poor OCT speckle reproduction.}
\label{fig:ssim_comparison}
\end{figure}

The generator used a `U-Net' architecture\cite{ronneberger_u-net:_2015} where a series of downsampling and upsampling convolutional paths with skip connections capture patterns at various levels of abstraction. A U-Net is traditionally used for image segmentation, but has also proven effective for the generation task. Recent deep learning papers have suggested that a deeper generator comprising multiple residual network (ResNet) blocks might have superior performance\cite{wang_high-resolution_2017}, but we did not observe significant differences with this on our training data. The generated image was fed to the discriminator, which was two `patchGAN'-style classifiers operating at two image scales\cite{isola_image--image_2016,wang_high-resolution_2017}. The receptive field of each pixel in the discriminator output was designed to be small (15 and 30 pixels width) relative to the input, such that finer details at the level of speckle might be evaluated by the discriminator. The GAN objective was regularized by an L1 (pixel-wise mean absolute difference) loss as follows: $L_{GAN}+\lambda L_{1}$ where $\lambda$ was a hyperparameter set to 10. Larger values of $\lambda$ up to 100 were suggested in prior studies using photographic data\cite{isola_image--image_2016}, but we found these to be prone to poor speckle generation and blurry images (Fig. \ref{fig:ssim_comparison}).  We also experimented with using an additional Difference of Structural Similarity (DSSIM) loss term as suggested in the literature\cite{zhao_loss_2017,wang_deep_2019} but this showed little improvement for OCT data and also produced blurry images; we have observed (Fig. \ref{fig:ssim_comparison}) that SSIM may be a poor training objective and evaluation metric for OCT generation. The Adam optimizer was used with learning rate 0.0001 and $\beta_1$=0.5, similar to previous reports\cite{isola_image--image_2016}. Batch size was 8 samples. For the first 2 epochs, the discriminator was trained on 1 step for every 3 steps trained by the generator; for the next 2 epochs, the discriminator was trained on 1 step for every 2 generator steps, and for subsequent epochs the discriminator and generator were trained equally. This is a commonly used heuristic for GAN training to prevent the generator from being overwhelmed by the discriminator too early, because the generator quality in the early epochs is expected to be much worse than real images. Other training heuristics previously recommended for GANs such as normalization of inputs between -1 and 1, soft and noisy labels were used\cite{chintala_github_2020}. Models were trained for 30 epochs with no early stopping or model pretraining/initialization. To further improve the quality of the images, Gaussian noise with standard deviation 0.1 was injected at every level of the generative upsampling, as well as at the input to the discriminator, as has been suggested by prior GAN reports\cite{karras_style-based_2018,shaham_singan:_2019,huszar_instance_2016}. Noise was injected in the generator during both the training and testing phase. Noise injection to the discriminator input has been experimentally shown in studies such as Huszar\cite{huszar_instance_2016} to increase the difficulty of the discriminator task and prevent the discriminator from overpowering the generator too quickly, especially during the early phases of training\cite{huszar_instance_2016}. The same model architecture and hyperparameters were used for enhancing both human lip and mouse skin data. Hyperparameters were not finely tuned, because objective evaluation of results was not possible due to the lack of robust quantitative metrics for quality, and thus no reasonable criterion existed for finely optimizing these hyperparameters. Even though model training was performed on image patches, the fully convolutional nature of the generator model (with no fully connected layers) allowed the use of image inputs that were larger than and not restricted to the training patch size of 256x256, therefore the full-sized original images could be used in the generator at model prediction time.

\subsection{Perceptual accuracy test by human OCT expert}

In order to evaluate the quality and realism of the computational reconstructions, a human reader was asked to evaluate the 2D-enhanced images. In typical GAN studies from the deep learning literature, human readers are crowdsourced from online platforms such as Amazon Mechanical Turk to evaluate the quality of generated photographs or artwork, but this is not feasible for specialized imaging data such as OCT. For our study, the reader was an OCT expert (coauthor X.L.) who was involved in planning the study and preparing the data, but was not involved in the machine learning, was blinded to the models, and had not seen the model-generated results beforehand. Image patches (256 x 256 pixels) were shown to the reader one at a time, and the reader given two seconds to evaluate. In a 'paired' test, a generated image was shown with its ground truth image side by side and the reader asked to identify the real image. In an 'unpaired' test, a single image, either generated or ground truth, was shown one at a time and the reader asked to determine its identity. Two seconds was longer than a typical GAN perceptual test, which gives only one second\cite{shaham_singan:_2019}, so as to account for the complexity of a typical OCT image. Before commencing the test, 5 practice examples were showed, each followed by the answer. This was then followed by a test of 50 questions in sequence, with no answers shown. After each test, a 'confusion score' was computed as the fraction of incorrectly read images over 50 total images, giving a percentage. Higher confusion scores closer to 50\% would indicate that many images were incorrectly read, suggesting that generated images were realistic and nearly indistinguishable from real high-resolution images (nearly random chance). Lower confusion scores closer to 0\% would indicate that most images were correctly read, suggesting that generated images were easily distinguished from real images. The confusion score of the GAN-generated results was compared to separate tests on images produced by a state-of-the-art Unet (non-adversarial training) originally designed for improving signal to noise ratio and image quality of microscopy images\cite{weigert_content-aware_2018}, as well as a vanilla conditional GAN\cite{isola_image--image_2016} without the additional injection of noise.

\subsection{Cross-domain validation on real data}

As a preliminary qualitative assessment of the models' performance on real (not simulated) images from a different data distribution from the training data, images of normal (no pathology) human retinal images\cite{gholami_octid_2019} and mouse bladder tissue\cite{zhou_optical_2019} were obtained from freely available datasets that accompanied published papers. Retinal images were acquired on a Zeiss Cirrus ophthalmic spectral domain OCT system, with axial and lateral resolution 5 $\mu$m and 15 $\mu$m respectively, and bladder tissue images were acquired on a Bioptigen Envisu R-class pre-clinical imaging system, with axial and lateral resolution 0.9 $\mu$m (tissue) and 8.5 $\mu$m respectively. It was necessary to resize the input images such that the size of speckle was approximately similar to that of the training data. The generator model, like most modern neural networks, used convolution operations, for which the convolutional filters had been learned from training data based on the length scales (measured in number of pixels) of image features including speckle noise. Therefore it was necessary for the images entering the generator to have roughly the same length scales of speckle learned by the generator's convolutional filters (Appendix). For datasets dissimilar to the training data, images were resized to 4x larger in pixel dimensions, using bilinear interpolation, before entering the generator. Low-signal regions deep in the images were cropped, and the images were marginally resized to have dimensions of a multiple of 256, for ease of input to the trained model. Since higher-resolution ground truths were not available, generated results were qualitatively assessed.

\begin{figure}[htbp]
\centering\includegraphics[width=10cm]{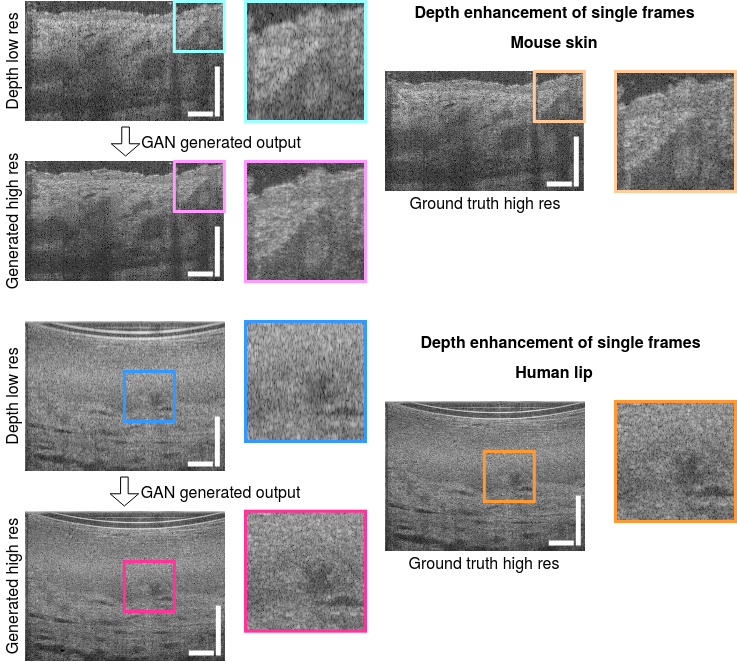}
\caption{Enhancement of depth resolution. Scale bar 100 $\mu$m.}
\label{fig:depLR}
\end{figure}

\section{Results and discussion}
We have developed a deep learning based algorithm for resolution enhancement of OCT images, based on previously reported techniques in generative adversarial networks. Using very high resolution OCT images as a ground truth, ~4x improvement in resolution was demonstrated on images with synthetic resolution degradation. As with typical GAN generation, objective evaluation of the generated outputs was challenging. Given the speckle noise that is inherent to coherent imaging such as OCT or ultrasound, the model was not able nor expected to exactly reproduce the noise content of the ground truth images. Therefore, conventional similarity metrics such as Structural Similarity (SSIM) gave low scores. Excessive regularization produced smoothed, speckle-reduced images with poor resemblance to OCT but still resulted in higher SSIM scores (Fig. \ref{fig:ssim_comparison}). Reduced regularization produced speckle noise that appeared qualitatively realistic, suggesting that the noise distribution of the speckle was learned, while the exact details of the generated speckle pattern was different from the ground truth. The generation of realistic yet accurate speckle may be necessary in some specific contexts, and is an interesting possibility for future investigation. Large regularization also seems to suggest itself as a means of speckle noise reduction, although this needs more careful validation and was not the objective of this work.

\begin{figure}[htbp]
\centering\includegraphics[width=10cm]{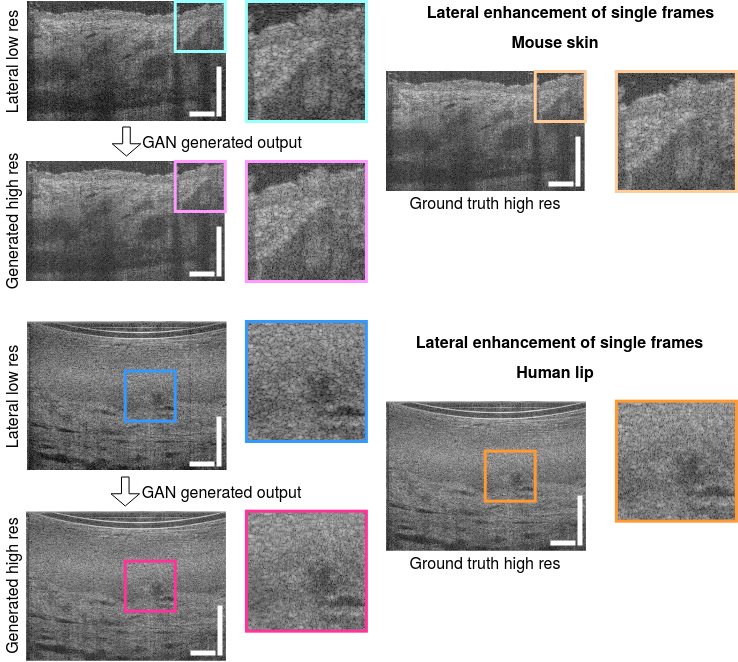}
\caption{Enhancement of lateral resolution. Scale bar 100 $\mu$m.}
\label{fig:latLR}
\end{figure}

\begin{figure}[htbp]
\centering\includegraphics[width=10cm]{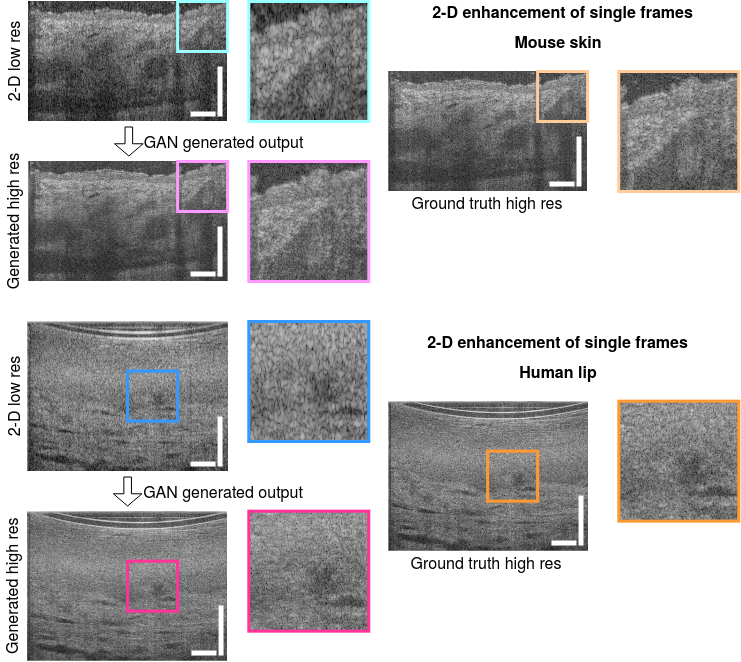}
\caption{Enhancement of depth and lateral (2-D) resolution. Scale bar 100 $\mu$m.}
\label{fig:2DLR}
\end{figure}

Image examples showing resolution enhancement in depth, lateral, and both (2-D) axes are shown in Figures \ref{fig:depLR}, \ref{fig:latLR}, \ref{fig:2DLR} and perceptual confusion scores presented in Table \ref{table:tab1}. The realism of generated speckle appeared to be high. However, it can be observed that the detailed content of the generated speckle pattern can differ from the ground truth, particularly in the 2-D case where the inference space is larger (Figure \ref{fig:2DLR}), even though the larger scale features are preserved and have improved quality. This difference in generated noise pattern, especially in the low-signal (nearly black) background of the images where noise can originate from the laser or other system sources, can lead to poor results when standard quantitative pixel-level  similarity metrics such as SSIM are used. 

\begin{table}[ht]
\centering
\begin{tabular}{|l|c c c c|} 
\hline
    \multirow{2}{*}{Data type and model} & \multicolumn{2}{c}{Single frame} & \multicolumn{2}{c|}{3-frame average} \\
     & Paired & Unpaired & Paired & Unpaired\\
% & \multicolumn{2}{c}{Single frame} & \multicolumn{2}{c|}{3-frame average} \\
% \hline
%Conversion & Human lip & Mouse skin & Human lip & Mouse skin\\  
 \hline
 Mouse skin Unet & 0 & 0 & 0 & 0\\ 
 \hline
 Mouse skin GAN\textsubscript{no noise} & 0  & 6  & 2 & 4\\
 \hline
 Mouse skin GAN\textsubscript{noise} & 42 & 22 & 52 & 26\\
 \hline  \hline
 Human lip Unet & 0 & 0 & 0 & 0\\ 
 \hline
 Human lip GAN\textsubscript{no noise} & 2 & 2 & 0 & 8\\ 
 \hline
 Human lip GAN\textsubscript{noise} & 4 & 6 & 20 & 22\\ 
 \hline

\end{tabular}
\caption{Confusion score in \% (0\%: zero confusion, $\sim$50\%: maximal confusion) from perceptual accuracy test by a human OCT expert reader on 2-D enhanced images, discriminating between a model output and ground truth. The test consisted of 50 images, such that confusion score was the fraction of incorrectly read images over 50 total images. Higher scores closer to 50\% indicate higher quality model outputs nearly indistinguishable from real high-resolution images (random chance).}
\label{table:tab1}
\end{table}

\begin{figure}[htbp]
\centering\includegraphics[width=10cm]{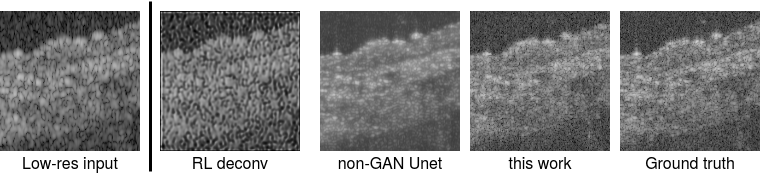}
\caption{Examples of images produced by a range of techniques. left-right: low-resolution input, Richardson-Lucy deconvolution with Gaussian point spread function of $\sigma=2$, non-adversarial Unet, noise-injected cGAN, ground truth.}
\label{fig:compare_techs}
\end{figure}

\begin{figure}[htbp]
\centering\includegraphics[width=7cm]{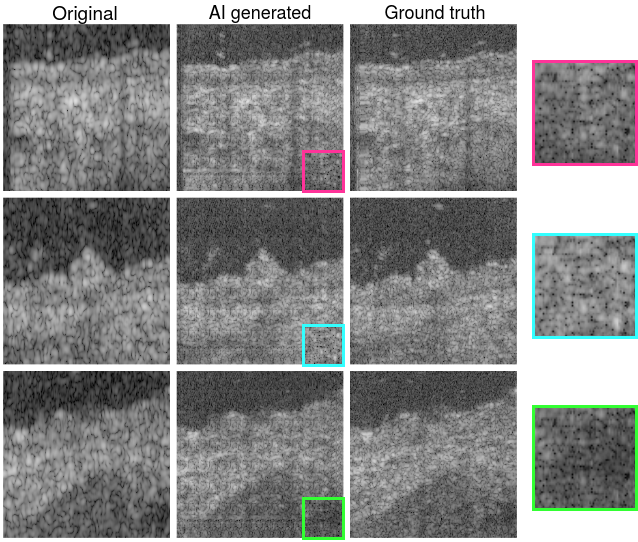}
\caption{Examples of failures from cGAN model with no injection of noise. Generated images have  repeated grid-like artifacts, and a repeated noise pattern (insets) in all images.}
\label{fig:failures}
\end{figure}

Human perceptual accuracy tests were preferred for evaluating the quality of the enhancement (Table \ref{table:tab1}). Examples from a range of algorithms are presented in Figures \ref{fig:compare_techs} and \ref{fig:failures}. The Richardson-Lucy technique was generally poor (Appendix) and thus deemed not sufficiently competitive for a human perceptual test. The non-adversarial Unet and vanilla cGAN (no noise injection) produced images that were easily discriminated by a human OCT expert (0\% confusion). The noise-injected cGAN confusion scores were substantially higher. The unpaired test results were lower than paired results, which was surprising and opposite to typical GAN studies\cite{shaham_singan:_2019} where readers found single images more confusing. This may be due to our reader being a subject-matter OCT expert, such that in the absence of a confusing alternate image, he was able to tap on pre-existing specialized knowledge of OCT to distinguish realistic speckle. Results from the 3-frame averaged images showed better quality (higher confusion). In practice, the interpretation of OCT images often involves an averaging/denoising process where speckle noise is intended to be suppressed; training a model on denoised data could allow the model to focus on more important image features rather than speckle noise, which is challenging to reproduce.  

\begin{figure}[htbp]
\centering\includegraphics[width=9cm]{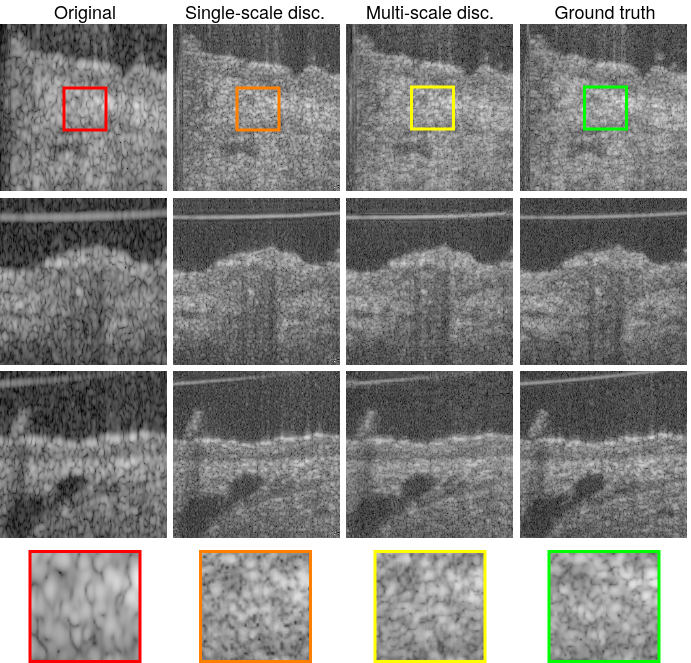}
\caption{Qualitative comparison of speckle generation by a single-scale and multi-scale discriminator, with insets for closer inspection. The latter produced speckles with slightly more variation in size, shape and intensity.}
\label{fig:disc_multiscale}
\end{figure}

The multi-scale discriminator resulted in more realistic speckle generation (Figure \ref{fig:disc_multiscale}). The single-scale discriminator produced speckles that have a chunky, artificial appearance, while the multi-scale discriminator produced speckles with slightly more variation in size, shape and intensity, although this can be subjective. 

Noise injection in the architecture was important for quality and realism of the reconstruction. Some examples of low quality images generated by a vanilla cGAN (no noise injection) are shown in Figure \ref{fig:failures}. The speckle pattern had a repeated grid-like artifact, severely affecting the realism of the images. The images also sometimes showed a noise pattern resembling speckle noise but repeated in most/all generated images (figure insets). This pattern might appear realistic on single images, but was quickly detected by the human reader as a generative artifact when observed over a large number of images from the same generator during a perceptual test.

The adversarial component of the algorithm appeared to be particularly important for OCT generation; the baseline non-adversarial Unet approach has been reported to be successful in microscopy for denoising, increasing signal-to-noise ratio and sharpening, but produced less realistic OCT images than our cGAN approach. Examples of images produced are shown in Figure \ref{fig:compare_techs}. This was in agreement with most computational super-resolution studies\cite{ledig_photo-realistic_2016,wang_super-resolution_2018}, which have favored GANs using adversarial learning. Our study had some important limitations. Small numbers of datasets were used in these proof-of-concept experiments, limiting the generalizability of the findings. Low-resolution training data was synthetically created, based on simple operations of spectral cropping and averaging, which may not sufficiently simulate low-resolution images from real-world conventional OCT systems. Real low-resolution images will be of even lower quality. While the use of spectral cropping to approximate low axial resolution may be closer to acceptable realism, the use of spectral averaging to approximate low lateral resolution is likely much less severe than the use of lower numerical aperture optics. The use of real-world low resolution data is an important future step towards rigorous validation of our system. Our network architecture mimicked a standard end-to-end learning design used in conventional GANs for artwork/photography, and did not incorporate physical or optical models of OCT image formation. Hybrid physics-inspired learning algorithms as suggested in recent computational optics studies\cite{monakhova_learned_2019} may potentially improve performance. The use of human perceptual accuracy tests, while advantageous for qualitatively evaluating OCT image quality and speckle restoration, should not be considered a rigorous test for resolution improvement. A single OCT expert reader was used to evaluate the images in the perceptual accuracy tests, therefore our results only demonstrate feasibility and may not generalize to other human experts. Future studies involving multiple readers will need to carefully control for specific levels of experience and familiarity with the specialized data, and successful recruitment may depend on availability of such experts.

\begin{figure}[htbp]
\centering\includegraphics[width=10cm]{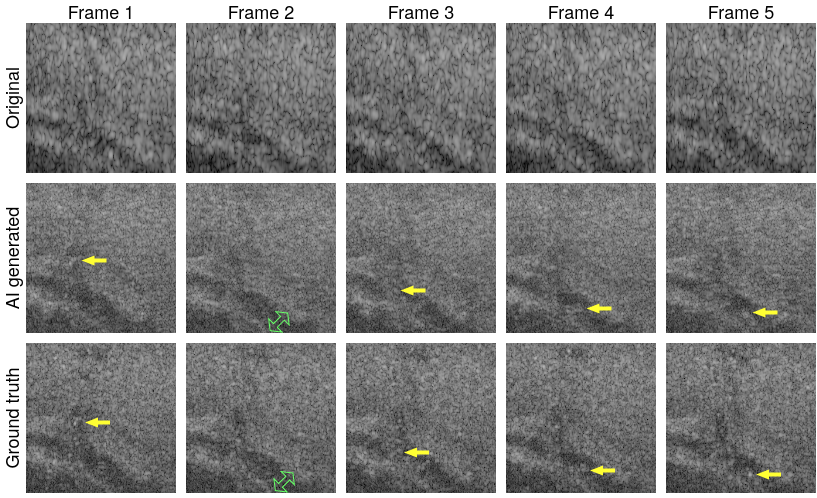}
\caption{Regions of interest cropped from lip images showing blood vessel (green arrow), across 5 consecutively acquired image frames from a single volumetric scan. Cellular particles (yellow arrows) flowing in the vessel cannot be reliably distinguished from surrounding speckles in the original low-resolution image, but are moderately restored by the AI resolution enhancement.}
\label{fig:vessel_particle}
\end{figure}

AI-based generative enhancement of resolution can have an important role in the use of high resolution OCT to study small biological features/particles at the cellular level. Fig. \ref{fig:vessel_particle} shows a series of lip images (cropped from a larger B-scan) acquired sequentially in a volumetric scan. The images show a blood vessel structure (green arrow), and each frame shows cellular particles (yellow arrows) flowing through the vessel. These particles can be seen with micro-OCT. At low resolution (top row), these particles are impossible to distinguish from speckle noise. The AI restoration process applied to low resolution images recovers the particles to a moderate extent, sufficiently to distinguish from the surrounding tissue speckles. Potentially, resolution enhancement based on AI tools could help in microscopic, dynamic analysis using conventional OCT imaging. 

\begin{figure}[htbp]
\centering\includegraphics[width=7cm]{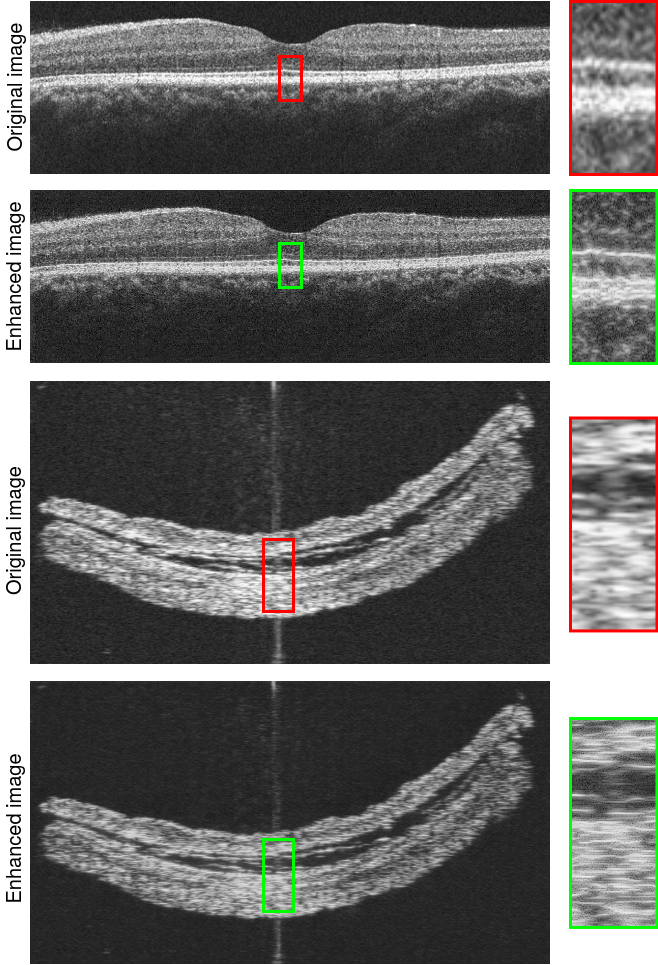}
\caption{Preliminary experiments with cross-domain application, applying a model trained on mouse skin micro-OCT to human retinal images (courtesy of \cite{gholami_octid_2019}) and mouse bladder images (courtesy of \cite{zhou_optical_2019}) from commercial OCT systems.}
\label{fig:domain_trans}
\end{figure}

As a preliminary step towards application on real data, publicly available retinal OCT images\cite{gholami_octid_2019} and mouse bladder OCT images\cite{zhou_optical_2019} were enhanced (Fig. \ref{fig:domain_trans}) using the model that performed best on perceptual tests (Table \ref{table:tab1}), the mouse skin model. Axial resolution was visibly enhanced, but lateral resolution enhancement was marginal. It should be noted that a simple deconvolution of these images with a carefully designed point spread function may have the same visual effect of enhanced axial resolution (thinner layers), so these experiments should be taken as illustrative in nature, and motivating of future work. In the future, a more realistic simulation of low lateral resolution in the training data (rather than simply a moving average of spectra) could further improve performance. In the retina, the resolution of layers particularly the inner/outer segments and retinal pigment epithelium (highlighted by inset in Figure \ref{fig:domain_trans}) was enhanced, which could have relevance to clinical thickness measurements that have been proposed in previous high-resolution OCT studies\cite{lu_photoreceptor_2017}. The model performance was found to be sensitive to the input image size; using the original image dimensions produced low quality results. We postulate this to be due to the size and scale of specific image features (e.g. speckle) that are learned from the training data (Appendix). The speckle size of input images should at least roughly match that of the training data; more robust protocols for domain transfer will be developed in future work. 

These early proof-of-concept experiments suggest the possibility of packaging the high performance of a prototype imaging system as a low cost software-based image enhancement tool that may be used by scientific/clinical peers who lack access to cutting-edge hardware. As long as the neural network model is trained on a data distribution that is identical or very similar to the intended test usage (e.g. imaging of the same organism and organ under similar conditions), the model can be expected to generate high-quality enhancement results. Even cross-domain OCT applications seem feasible in principle, based on our preliminary experiments, although results will be more variable and will require careful validation. Concerns of super-resolution inference leading to 'hallucinatory' artifacts have been reported\cite{cohen_distribution_2018}, which should not dampen enthusiasm for this research direction but motivate validation by human readers and comparisons with ground truth images. This concept may also be relevant to high speed swept source OCT systems\cite{liang_endoscopic_2017} that typically have lower optical bandwidth and thus worse axial resolution than spectral domain systems. The possibility of having 'the best of both worlds' of OCT systems combining high speed and high resolution is an intriguing avenue of future investigation.

\section{Conclusion}

In this proof-of-concept study, the feasibility of axial and lateral resolution enhancement of OCT images using a generative adversarial network was investigated. A high resolution ground truth acquired with micro-OCT, paired with simulated low resolution image inputs were used to train a neural network to generate resolution-enhanced outputs. Results were evaluated by a human OCT expert for perceptual realism. Preliminary cross-domain experiments were performed on image data from outside of the training data distribution. Future work will involve the acquisition of more realistic training data, such as true low lateral resolution images taken with low numerical aperture optics and low axial resolution images taken with reduced source bandwidth, larger amounts of data with more variety of quality including typical imaging artifacts, and studies of cross-domain transferability and robustness.

\section*{Funding}
This research is supported by the Ministry of Education Singapore under its Academic Research Fund Tier 1 (2018-T1-001-144), National Medical Research Council Singapore under its Open Fund - Individual Research Grant (MOH-OFIRG19may-0009), and Agency for Science, Technology and Research (A*STAR) under its Industrial Alignment Fund (Pre-positioning) (H17/01/a0/008).

\section*{Acknowledgments}
We thank Dr. Chen-Hsin Sun for helpful conversations on ophthalmic OCT.

\section*{Disclosures}
The authors declare no conflicts of interest.

%%%%%%%%%%%%%%%%%%%%%%% References %%%%%%%%%%%%%%%%%%%%%%%%%

%%%%%%%%%% If using BibTeX:
\bibliography{ai-oct-paper}

\newpage

\section{Appendix}

\subsection{Image resizing for cross-domain transfer of generative models}

\begin{figure}[htbp]
\centering\includegraphics[width=6cm]{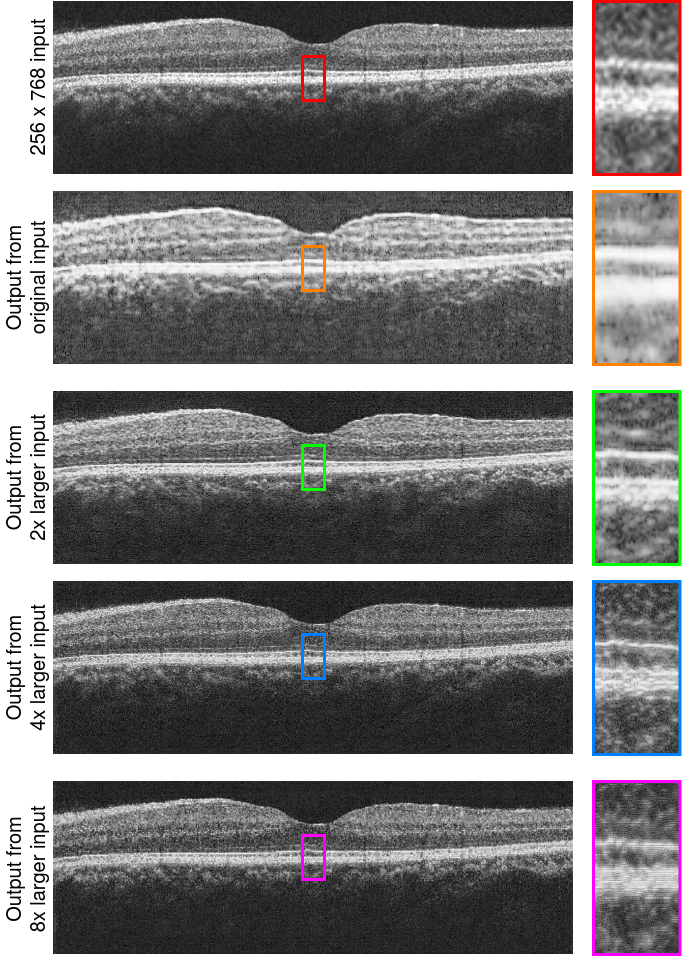}
\caption{Generated images from dataset (retina) outside of the training data distribution (mouse skin), with various scale factors on the input image.}
\label{fig:eye_super}
\end{figure}

As described in Methods, the input image to the generator should have length scale of features approximately matched to that of the generator's convolutional filters (learned from training data). Figure \ref{fig:eye_super} shows the effect of image scaling. Beyond 4x on input scaling, the generator appeared to produce unrealistic artifacts. In future studies of cross-domain generative transfer, the scale factor can be a hyperparameter to be optimized, although generated images will still require careful inspection by human experts for quality and realism.

\subsection{Richardson-Lucy deconvolution parameter selection}

\begin{figure}[htbp]
\centering\includegraphics[width=7cm]{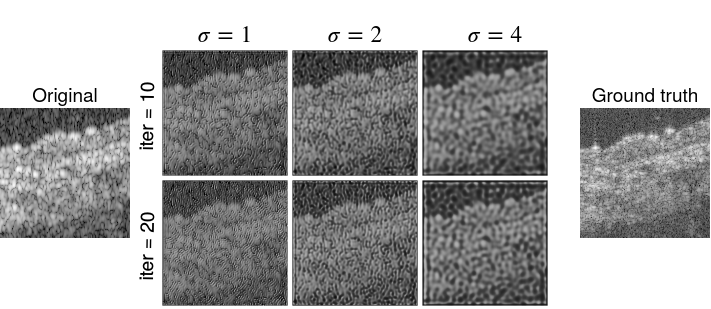}
\caption{Images from Richardson-Lucy deconvolution with various parameters.}
\label{fig:rl_params}
\end{figure}

Richardson-Lucy deconvolution was generally unable to produce good quality or realistic OCT images of higher resolution. The point spread function was estimated by a Gaussian. The generated images were sensitive to the parameters of $\sigma$, the standard deviation of the Gaussian PSF, and number of iterations. Some examples are shown in Figure \ref{fig:rl_params}. 

\subsection{Quantitative metrics for image enhancement}

While quantitative metrics are not able to adequately capture improvements in image content and quality (Table \ref{table:tab2}), they are still important evaluations that complement perceptual tests. The low metric scores indicate that the enhanced images are still substantially different from the ground truth images at a pixel-level comparison, which is a downside to these generative approaches. Speckle noise is difficult to be recovered deterministically at the pixel-level. In many scenarios, pixel-level differences may be less important than the recovery of high-resolution content that could indicate important tissue features.

\begin{figure}[htbp]
\centering\includegraphics[width=8cm]{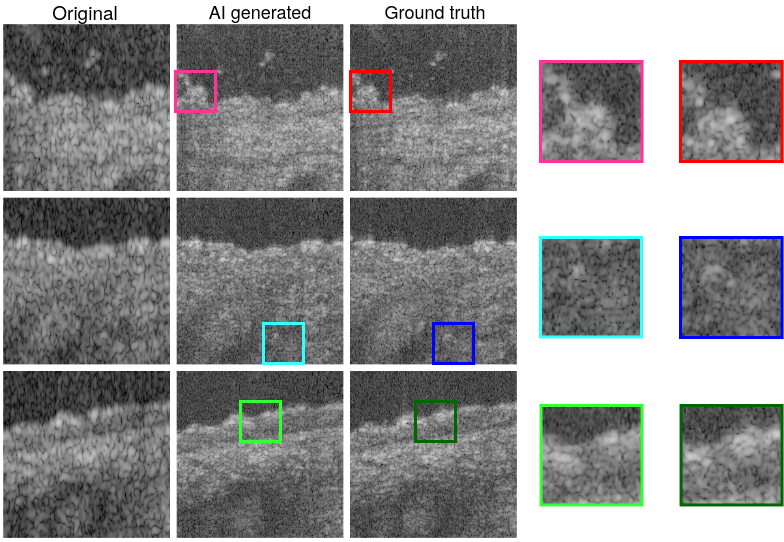}
\caption{Qualitative comparison of generated patches and ground truths, illustrating good agreement of image features on a global scale but significant pixel-level deviations.}
\label{fig:realfake_compare}
\end{figure}

The low quantitative metrics are illustrated in Figure \ref{fig:realfake_compare}, where generated patches appear accurate on a global scale and of good quality, but closer inspection (insets) reveal pixel-level deviations from the ground truth. Generally OCT interpretation is qualitative and does not rely on detailed analysis on individual speckles, but in more demanding applications where speckles are measured or quantified, such deviations would be of greater concern. 

\begin{table}[ht]
\centering
\begin{tabular}{|l|c c c c|} 
\hline
    \multirow{2}{*}{Enhance w/ noise injection} & \multicolumn{2}{c}{Single frame} & \multicolumn{2}{c|}{3-frame average} \\
     & SSIM & PSNR & SSIM & PSNR\\
% & \multicolumn{2}{c}{Single frame} & \multicolumn{2}{c|}{3-frame average} \\
% \hline
%Conversion & Human lip & Mouse skin & Human lip & Mouse skin\\  
 \hline
 Mouse skin depth enhance & 0.295$\rightarrow$0.263 & 21.4$\rightarrow$21.6 & 0.318$\rightarrow$0.249 & 24.9$\rightarrow$25.0\\ 
 \hline
 Mouse skin lateral enhance & 0.217$\rightarrow$0.288  & 20.3$\rightarrow$21.7  & 0.282$\rightarrow$0.333 & 23.0$\rightarrow$26.1\\
 \hline
 Mouse skin 2-D enhance & 0.085$\rightarrow$0.104 & 19.9$\rightarrow$20.5 & 0.122$\rightarrow$0.131 & 22.9$\rightarrow$24.2\\
 \hline  \hline
 Human lip depth enhance & 0.382$\rightarrow$0.347 & 21.1$\rightarrow$21.9 & 0.393$\rightarrow$0.394 &  23.5$\rightarrow$26.2\\ 
 \hline
 Human lip lateral enhance & 0.238$\rightarrow$0.262 & 20.3$\rightarrow$21.8 & 0.291$\rightarrow$0.352 & 23.0$\rightarrow$26.6\\ 
 \hline
 Human lip 2-D enhance & 0.096$\rightarrow$0.125  & 19.8$\rightarrow$20.7 & 0.126$\rightarrow$0.165 & 22.4$\rightarrow$23.8\\ 
 \hline

\end{tabular}

\begin{tabular}{|l|c c c c|} 
\hline
    \multirow{2}{*}{Enhance w/o noise injection} & \multicolumn{2}{c}{Single frame} & \multicolumn{2}{c|}{3-frame average} \\
     & SSIM & PSNR & SSIM & PSNR\\
% & \multicolumn{2}{c}{Single frame} & \multicolumn{2}{c|}{3-frame average} \\
% \hline
%Conversion & Human lip & Mouse skin & Human lip & Mouse skin\\  
 \hline
 Mouse skin depth enhance & 0.295$\rightarrow$0.184 & 21.4$\rightarrow$20.5 & 0.318$\rightarrow$0.285 & 24.9$\rightarrow$27.2\\ 
 \hline
 Mouse skin lateral enhance & 0.217$\rightarrow$0.184  & 20.3$\rightarrow$21.0  & 0.282$\rightarrow$0.265 & 23.0$\rightarrow$25.3\\
 \hline
 Mouse skin 2-D enhance & 0.085$\rightarrow$0.073 & 19.9$\rightarrow$20.3 & 0.122$\rightarrow$0.114 & 22.9$\rightarrow$24.1\\
 \hline  \hline
 Human lip depth enhance & 0.382$\rightarrow$0.241 & 21.1$\rightarrow$21.2 & 0.393$\rightarrow$0.289 &  23.5$\rightarrow$22.0\\ 
 \hline
 Human lip lateral enhance & 0.238$\rightarrow$0.208 & 20.3$\rightarrow$21.6 & 0.291$\rightarrow$0.266 & 23.0$\rightarrow$25.2\\ 
 \hline
 Human lip 2-D enhance & 0.096$\rightarrow$0.085  & 19.8$\rightarrow$20.4 & 0.126$\rightarrow$0.149 & 22.4$\rightarrow$24.0\\ 
 \hline

\end{tabular}

\caption{Pixel-level quantitative metrics, including Structural Similarity (SSIM) and Peak Signal to Noise Ratio (PSNR), comparing ground truth and before/after AI-based enhancement, and with/without noise injection. These metrics do not adequately reflect improvements in quality and realism, but indicate that generated images deviate from ground truth significantly at pixel level.}
\label{table:tab2}
\end{table}

\end{document}